\documentclass[11pt,a4paper,logo]{xiaomi}
\usepackage[authoryear,sort&compress,round]{natbib}
\usepackage{amsmath,amsfonts,bm}

\def\eqref#1{equation~\ref{#1}}

\def\1{\bm{1}}

\DeclareMathAlphabet{\mathsfit}{\encodingdefault}{\sfdefault}{m}{sl}
\SetMathAlphabet{\mathsfit}{bold}{\encodingdefault}{\sfdefault}{bx}{n}

\hypersetup{hidelinks}
\usepackage{url}
\usepackage[normalem]{ulem}
\usepackage{color}
\usepackage{textgreek}
\usepackage{multirow}
\usepackage{algorithmicx,algorithm}
\usepackage[noend]{algpseudocode}
\usepackage{subfig}
\usepackage{array}
\usepackage{colortbl}
\usepackage{enumitem}
\usepackage{xspace}
\usepackage{amsmath}
\usepackage{amssymb}
\usepackage{booktabs}
\usepackage{tabularx}
\usepackage{graphicx}
\usepackage{caption}
\usepackage{pifont}
\usepackage{etoolbox}
\usepackage{tikz}
\usetikzlibrary{positioning}

\setlist{nolistsep}

\newcommand{\cut}[1]{}

\newcommand{\sysname}{\textsc{Prorsa}\xspace}

\title{\sysname: Unleashing the Potential of Prefill-Decode Disaggregation in Agentic Reinforcement Learning}

\DeclareRobustCommand{\paperauthors}{
\vskip5pt
{\authorfont Chiheng Lou$^{\dagger*}$ ~~~~~Xiangwei Deng$^{\ddagger}$ ~~~~~Bangjun Xiao$^{\dagger*}$ ~~~~~Xiaodong Ji$^{\ddagger}$} \\
{\authorfont Yikai Zhao$^{\dagger\ddagger}$ ~~~~~Xin Jin$^{\dagger}$ ~~~~~Hailin Zhang$^{\ddagger\diamond}$} \\
\vskip10pt
{\authorfont $^{\dagger}$School of Computer Science, Peking University} \\
{\authorfont $^{\ddagger}$LLM-Core Xiaomi}
\vskip10pt
}
\author{\paperauthors}

\begin{abstract}
\label{sec:abstract}

Agentic reinforcement learning (RL) is commonly used to enhance the capabilities of large language models (LLMs) in solving complex multi-turn tasks.
During rollout, agentic RL exhibits a workload pattern distinct from that of traditional RL: multi-turn interactions repeatedly introduce prefill requests, causing prefill and decode workloads to coexist throughout rollout.
This workload makes prefill-decode (PD) disaggregation, a design widely used for online LLM serving, promising for speeding up agentic RL rollout.
This paper shows that PD disaggregation can improve rollout efficiency by (a) mitigating interference among data-parallel attention modules, (b) enabling stage-specific parallelism, and (c) reducing end-to-end trajectory completion time.
However, a fixed prefill-decode allocation may fail to sustain these benefits as workload composition varies across tasks and over time.
We therefore present \sysname, an agentic RL framework for prediction-guided prefill-decode resource allocation.
\sysname observes the lifecycle of every trajectory to forecast short-horizon prefill and decode demand.
These forecasts then guide worker allocation decisions and trigger role switching of rollout engines,
aligning prefill and decode capacity with anticipated workload demand.
Across synchronous and asynchronous agentic RL settings, \sysname achieves end-to-end rollout speedups of 1.11$\times$--4.05$\times$ over prefill-decode colocation.
\end{abstract}

\begin{document}

\maketitle
\begingroup
\renewcommand{\thefootnote}{\fnsymbol{footnote}}
\footnotetext[0]{$^{*}$Work done during internship at Xiaomi Corporation.}
\footnotetext[0]{$^{\diamond}$Corresponding author.}
\endgroup

\section{Introduction}
\label{sec:introduction}

Reinforcement learning (RL) has become a core stage of post-training for large language models (LLMs)~\citep{mai_thinking_1, deepseek-r1, qwen3technicalreport, glm5, deepseekai2026deepseekv4highlyefficientmilliontoken, kimiteam2026kimik3openfrontier}.
Beyond single-turn reasoning, RL is increasingly applied to \emph{agentic} scenarios---coding agents~\citep{pan2024swe-gym,wei2025swerladvancingllmreasoning}, computer use~\citep{wang2025uitars2technicalreportadvancing,huang2026evocua15onlinereinforcementlearning,lu2026qwencuanativecomputeruse}, and deep research~\citep{jin2025searchr1trainingllmsreason,zheng2025deepresearcherscalingdeepresearch}---%
where the model accomplishes a long-horizon task through multiple turns of interaction with the environment.

During RL training, the rollout phase generates samples from prompts~\citep{sheng2025hybridflow,hu2024openrlhf}.
For agentic RL, a sample is no longer a single model response but a long-lived trajectory that interleaves LLM generation with tool execution across many turns.
In this case, rollout can account for up to 70\% of end-to-end iteration time, making its efficiency an important optimization target~\citep{zhong2024rlhfuse, fu2025areal, gao2025rollpacker,heddle,chen2026libraefficientresourcemanagement}.

This paper studies how to improve agentic rollout efficiency through prefill-decode (PD) disaggregation, which places the two inference stages on separate GPU groups.
The benefits of PD disaggregation have been well studied in online LLM serving,
where it enables independent resource allocation and reduces interference between the two stages~\citep{distserve, splitwise, mooncake, hu2024inference, hong2025semi, feng2025windserve, zhu2025nanoflow}.
Its benefits in agentic RL rollout, however, have not been clearly characterized by prior works~\citep{glm5,minimaxm2,rollart}.

Our analysis shows that agentic RL exhibits distinct characteristics that make PD disaggregation particularly beneficial.
Specifically, agentic RL generates a continuous stream of prefill requests during rollout.
The prefill workloads come from two sources:
(1) \emph{multi-turn interactions}, where each trajectory repeatedly issues inference requests across interaction turns, creating prefill demand as new environment feedback is appended to its context,
and (2) \emph{asynchronous rollout}, where new samples are continuously injected into rollout engines rather than all at once to mitigate the long-tail problem~\citep{fu2025areal, zhong2025streamrl, han2025asyncflow, zhou2025aprilactivepartialrollouts}.
As a result, prefill and decode requests coexist throughout the rollout phase~(\S\ref{sec:workload-charac}).

For these interleaved rollout workloads, we find three key benefits of PD disaggregation: (1) reducing GPU idleness caused by prefill-decode interference across data-parallel attention modules, (2) enabling independent parallelism strategies for prefill and decode to maximize throughput, and (3) reducing decode latency for long-tail trajectories.

However, applying PD disaggregation to agentic RL is not straightforward.
Prefill and decode workloads vary across RL tasks, complicating the choice of a suitable prefill-to-decode instance ratio (PD ratio).
Moreover, the prefill and decode demands change over time even within the same task, so a fixed PD ratio may fail to sustain a throughput advantage over colocation throughout training.

To address this challenge, we propose \sysname, an agentic RL framework that combines PD disaggregation with adaptive worker allocation to improve rollout efficiency.
The key insight of \sysname is that the RL framework's visibility into trajectory lifecycles provides advance signals of upcoming inference demand.
By monitoring sample dispatch, sandbox startup, tool execution, and reward completion, \sysname forecasts short-horizon prefill and decode demand.
Based on these forecasts, \sysname adjusts the PD ratio by switching rollout workers between prefill and decode roles.
This adjustment mechanism keeps the instance allocation aligned with changing workload demands, improving GPU utilization.

Our main contributions are as follows:
\begin{itemize}[leftmargin=*]
    \item We characterize the benefits and deployment challenges of PD disaggregation in agentic RL rollout through an in-depth workload and performance analysis.
    \item We introduce a prefill-decode role switching strategy driven by short-horizon workload forecasts to adapt resource allocation to changing rollout demand.
    \item We design and implement \sysname, an agentic RL framework that adopts PD disaggregation to improve rollout efficiency. Experimental results across various agentic RL scenarios demonstrate that \sysname achieves rollout speedups ranging from 1.11$\times$ to 4.05$\times$ over colocation in both synchronous and asynchronous settings.
\end{itemize}
\section{Background}
\label{sec:background}

\subsection{Agentic RL Training}
\label{sec:rl-background}

In RL training, each iteration includes a \textit{rollout} phase that generates trajectories using inference engines~\citep{vllm,sglang}, a \textit{reward} phase that scores the trajectories with predefined rules or reward models, and a
\textit{training} phase that updates the model with an RL algorithm such as PPO~\citep{schulman2017proximal}, GRPO~\citep{shao2024deepseekmath}, or DAPO~\citep{yu2025dapo}.
Training begins once enough trajectories are scored,
and updated weights are synchronized to rollout engines.

In agentic RL, a sample is no longer a single response but a multi-turn trajectory:
in each turn the model emits an action (e.g., a tool call),
the environment executes it, and the feedback is appended to the context for the next turn.
For on-policy training, the rollout is \emph{synchronous}:
training cannot begin until all trajectories in a batch are complete, so long-tail trajectories leave GPUs idle.
To mitigate this effect, large-scale systems often adopt \emph{asynchronous} rollout~\citep{mai_thinking_1,deepseekai2026deepseekv4highlyefficientmilliontoken,glm5,fu2025areal, zhong2025streamrl, han2025asyncflow}, which continuously injects new samples and allows off-policy updates using trajectories generated by earlier model versions.
In this case, a \emph{staleness} bound limits the number of model versions a trajectory can span, and trajectories that fail to finish within this bound are dropped.

\subsection{Prefill-Decode Disaggregation}
\label{sec:pd-disaggregation}

LLM inference consists of two stages:
the \emph{prefill} stage processes all prompt tokens in parallel and produces the first output token,
while the \emph{decode} stage generates the remaining tokens autoregressively.
Prefill-decode disaggregation~\citep{distserve} places the two stages on separate instances:
prefill instances process prompts and transfer the produced \emph{KV cache} to decode instances, which carry out generation.
In online serving, disaggregation prevents decode from being blocked by prefill, and enables separate configuration of parallelism, accelerator types,
and instance counts for prefill and decode to meet their respective latency SLOs~\citep{distserve, splitwise}.
{}
\section{Motivation and Challenges}
\label{sec:workload}

In this section, we first characterize agentic RL workloads (\S\ref{sec:workload-charac}), and then show the benefits of PD disaggregation (\S\ref{sec:motivation}) under continuous prefill demand.
Finally, we present the challenges of adopting PD disaggregation for dynamic RL rollout workloads (\S\ref{sec:challenge}).

\subsection{Workload Characterization of Agentic RL Rollout}
\label{sec:workload-charac}

\begin{figure}[t]
\centering
\includegraphics[width=0.95\linewidth]{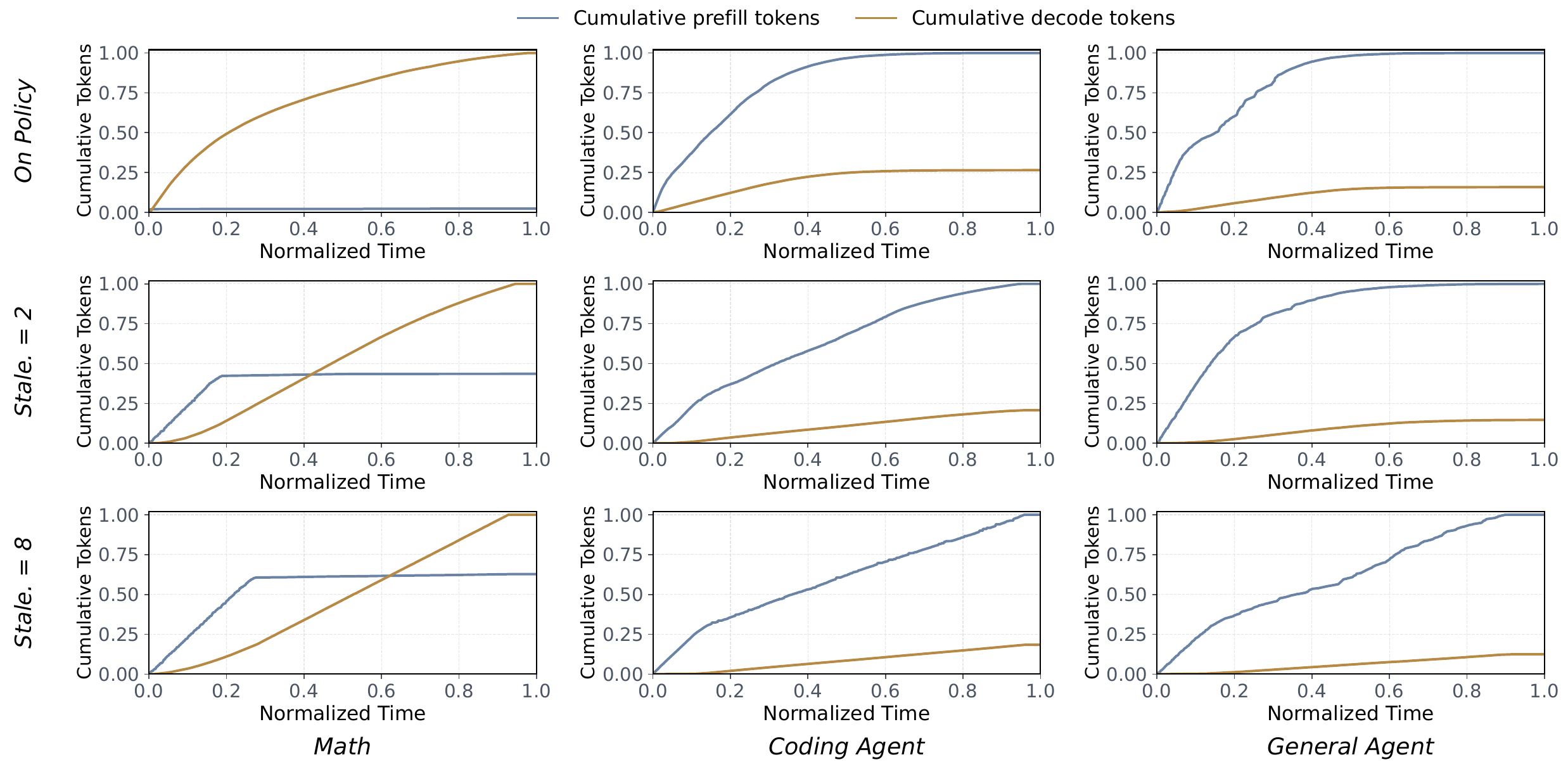}
\caption{Normalized cumulative prefill and decode tokens for different RL tasks.}
\label{fig:workload-grid}
\end{figure}

Figure~\ref{fig:workload-grid} shows the cumulative prefill and decode tokens within one rollout step for three representative tasks: math, coding agent, and general agent. The details of the tasks are described in \S\ref{sec:eval-setup}.
We vary the maximum staleness to capture the effects of asynchronous rollout.
At the start of each step, trajectories carried over from the previous step must re-prefill their existing contexts with the latest model version~\citep{fu2025areal,mimo2026v25pro}.

The tasks exhibit distinct workload patterns.
For math, prefill is concentrated near the beginning of each step, primarily due to initial prompt processing and, in asynchronous settings, re-prefill of trajectories carried over from the previous step.
In contrast, the two agentic tasks generate prefill requests throughout the step, and their prefill-token volumes substantially exceed their decode-token volumes.
Under the on-policy setting, the token-processing rate drops sharply near the end of the step, revealing a long tail of straggling trajectories~\citep{zhong2024rlhfuse,zhong2025streamrl,gao2025rollpacker,heddle,zhou2025aprilactivepartialrollouts}.

\subsection{Motivating Experiments}
\label{sec:motivation}

\begin{figure}[t]
  \centering
  \begin{minipage}[t]{0.48\linewidth}%
    \vspace{0pt}%
    \captionof{table}{Normalized per-server prefill and decode throughput for MiMo-V2.5-Pro. Decode uses a batch size of 16 per attention DP rank under all configurations.}%
    \label{tab:pd-throughput-ep}%
    \centering
    \footnotesize
    \setlength{\tabcolsep}{7pt}%
    \begin{tabular*}{0.95\linewidth}{@{\extracolsep{\fill}}ccccc@{}}
      \toprule
      Stage & Length & EP16 & EP32 & EP64 \\
      \midrule
      \multirow{4}{*}{Prefill}
        & 8K  & \textbf{3.24} & 2.78 & 2.22 \\
        & 64K & \textbf{2.77} & 2.33 & 1.95 \\
        & 256K & \textbf{1.76} & 1.36 & 1.37 \\
        & 512K & \textbf{1.29} & 1.02 & 1.00 \\
      \midrule
      \multirow{4}{*}{Decode}
        & 8K  & 1.20 & \textbf{1.31} & 1.29 \\
        & 64K & 1.17 & \textbf{1.26} & 1.25 \\
        & 256K & 1.10 & \textbf{1.14} & 1.13 \\
        & 512K & \textbf{1.02} & 1.01 & 1.00 \\
      \bottomrule
    \end{tabular*}%
  \end{minipage}%
  \hfill
  \begin{minipage}[t]{0.48\linewidth}%
    \vspace{0pt}%
    \centering
    \includegraphics[width=\linewidth]{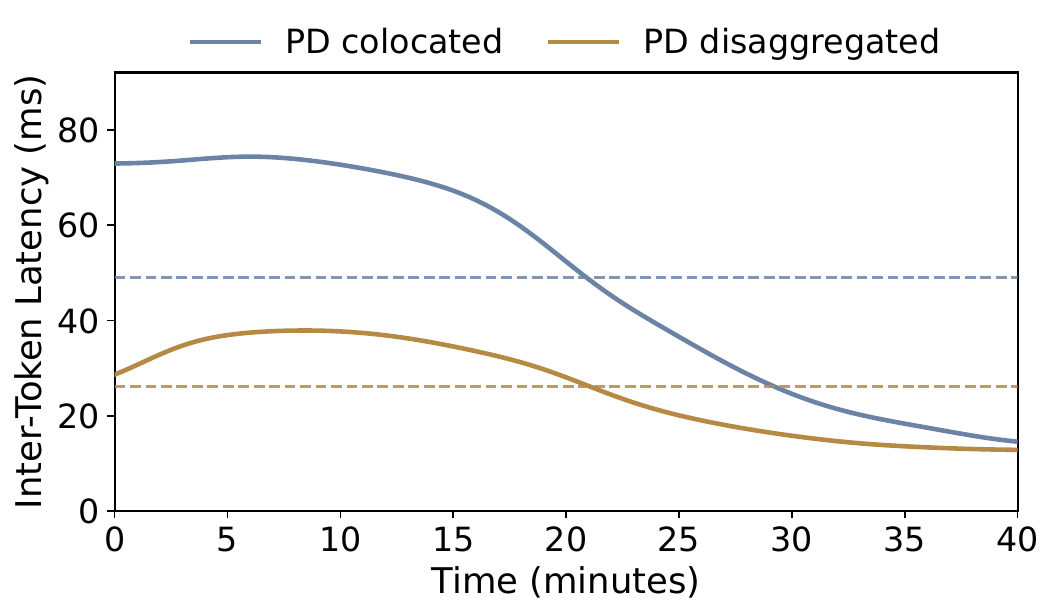}%
    \captionof{figure}{Inter-token latency of PD disaggregation versus colocation for synchronous rollout on the coding agent task. Dashed lines indicate the corresponding mean latencies.}%
    \label{fig:motivation-itl}%
  \end{minipage}%
  \vspace{-1em}
\end{figure}

\begin{figure}[t]
  \centering
  \includegraphics[width=\linewidth]{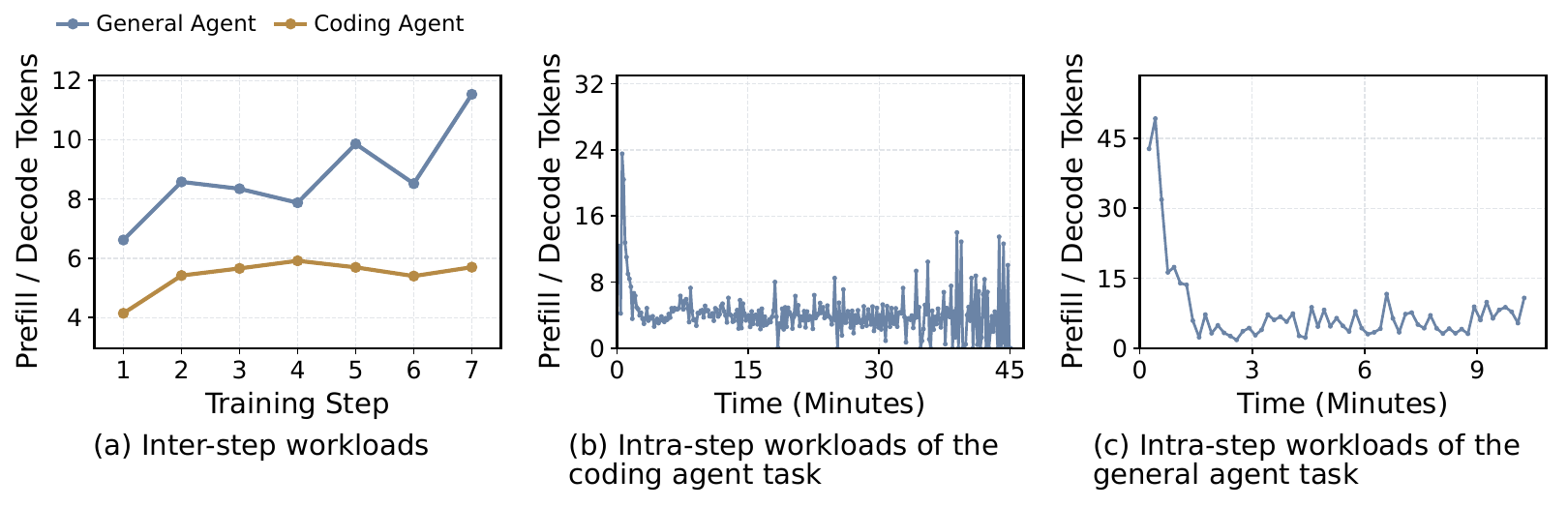}
  \caption{Prefill-to-decode token ratios across and within steps with a maximum staleness of 8.}
  \label{fig:workload-cross-step}
\vspace{-1em}
\end{figure}

Given this continuous prefill demand in agentic RL rollout, this paper shows that PD disaggregation can improve rollout efficiency.
Using MiMo-V2.5-Pro~\citep{mimo2026v25pro} and the experimental setup in \S\ref{sec:eval-setup}, we identify three major sources of performance improvement:

\textbf{Mitigating interference among data-parallel attention modules.}
Prior work has studied how PD colocation increases \emph{decoding latency} through head-of-line blocking~\citep{distserve,splitwise}.
However, we show that PD colocation can also significantly degrade \emph{throughput} for mixture-of-experts (MoE) models.
In MoE models, experts account for most model parameters, so their deployment commonly combines large-scale expert parallelism with multiple data-parallel (DP) attention modules~\citep{deepseek-v3,mimo2026v25pro,glm5}.

In this case, when prefill and decode are colocated, their differing execution times cause interference across attention DP ranks:
if one rank processes prefill while others process decode, the decoding ranks must wait before entering the synchronized MoE phase, leaving GPUs idle.
Chunked prefill~\citep{agrawal2023sarathi} shortens each stall but cannot eliminate the duration mismatch.
This synchronization-induced idleness can reduce rollout throughput:
for MiMo-V2.5-Pro on the two agentic tasks, we observe a GPU idle ratio of up to 26\% due to prefill-decode interference (\S\ref{sec:perf-breakdown}).

\textbf{Independent parallelism strategies for prefill and decode.}
PD disaggregation also allows each stage to use its own parallelism strategy~\citep{distserve,forys2026doesdisaggregationpaysimulating}.
Table~\ref{tab:pd-throughput-ep} shows the normalized per-server throughput of prefill and decode under different parallelism strategies and input lengths using the testbed in \S\ref{sec:eval-setup}.
The results show that prefill consistently favors EP16, whereas decode favors EP32 for request lengths up to 256K and EP16 at 512K.
Although EP64 does not maximize decode throughput, it provides more KV-cache capacity, which can be necessary for extremely long requests.
These measurements motivate selecting parallelism separately for prefill and decode instances.

\textbf{Reduced decode latency.}
RL rollout time is often dominated by a small number of long-tail trajectories, and prior work primarily addresses them by reducing decode latency~\citep{zhong2024rlhfuse,zhong2025streamrl,gao2025rollpacker,heddle}.
PD disaggregation reduces decode latency by isolating decode from prefill interference.
For synchronous rollout on the coding agent task, Figure~\ref{fig:motivation-itl} shows that PD disaggregation achieves significantly lower inter-token latency than colocation.
The reduced decoding latency can shorten long-running trajectories, thus alleviating the long-tail problem.

\subsection{Challenges}
\label{sec:challenge}

The benefits of PD disaggregation depend on assigning the right number of workers to each stage.
However, as shown in Figure~\ref{fig:workload-grid}, workload composition varies substantially across tasks, requiring task-specific PD allocations.
Moreover, this composition also varies during training.
Figure~\ref{fig:workload-cross-step}(a) shows the prefill-to-decode token ratios across training steps with a maximum staleness of 8.
The maximum-to-minimum ratio can reach 1.74$\times$, as asynchronous sample dispatch leads to additional variability in rollout demand.
Figure~\ref{fig:workload-cross-step}(b) and Figure~\ref{fig:workload-cross-step}(c) further show sharp workload fluctuations within the first training step at 10-second granularity.
The complete workload analysis across different staleness settings is provided in Appendix~\ref{sec:appendix-motivation}.
Together, these variations motivate dynamically adjusting the PD worker ratio rather than maintaining a fixed allocation throughout training.
\section{\sysname Design}
\label{sec:design}

\begin{figure}[t]
  \centering
  \includegraphics[page=1,width=0.95\linewidth,keepaspectratio]{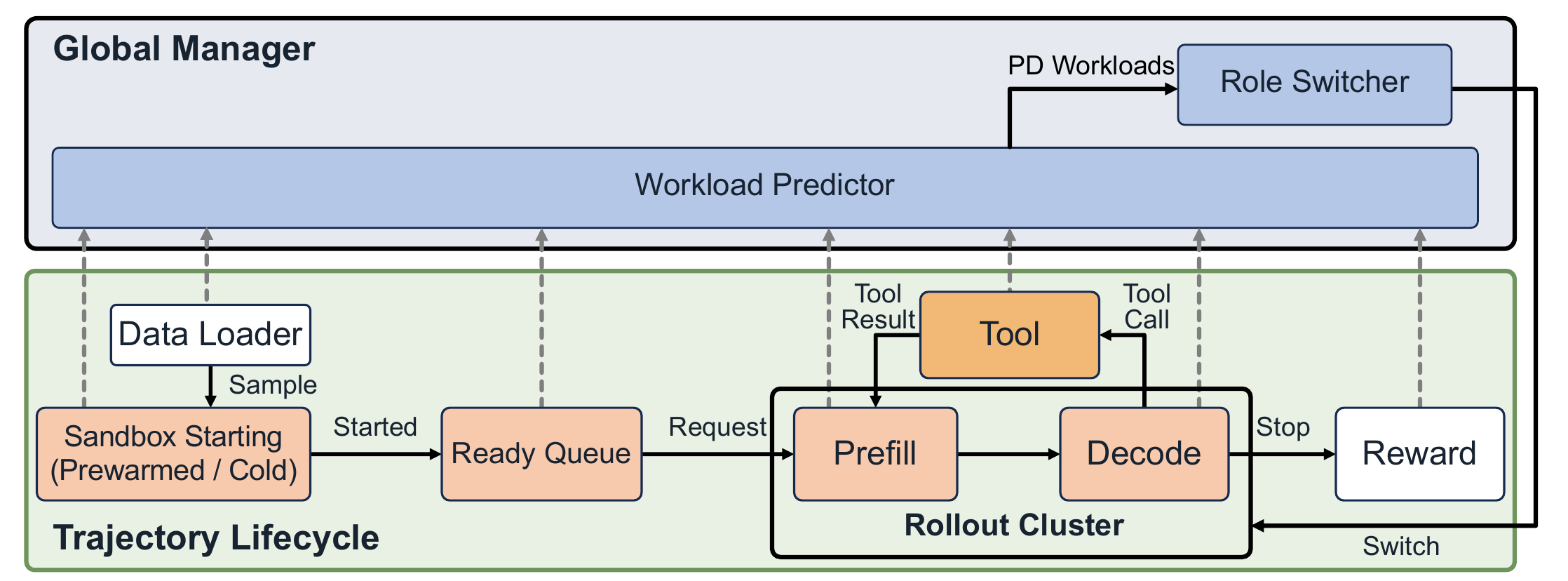}
  \caption{\sysname architecture overview.}
  \label{fig:overview}
\vspace{-1em}
\end{figure}

To adapt prefill-decode disaggregation to dynamic rollout workloads, we propose \sysname, an agentic RL framework with prediction-guided worker allocation.
Figure~\ref{fig:overview} presents the architecture of \sysname.
The framework leverages its visibility into each trajectory's lifecycle, including sample dispatch, sandbox startup, inference requests, tool completion, and reward computation.
Using these lifecycle signals, \sysname forecasts short-horizon prefill demand and decode KV-cache requirements (\S\ref{sec:design-prediction}) and dynamically switches rollout workers between the two roles (\S\ref{sec:design-switch}) to accommodate upcoming requests.

\subsection{Workload Predictor}
\label{sec:design-prediction}

\subsubsection{Trajectory State Machine}
To forecast future demands, \sysname traces each trajectory from sampling to final reward.
Figure~\ref{fig:overview} shows the state machine of every trajectory.
Each sampled trajectory starts a sandbox, either from the prewarm pool or through a cold startup.
Once the sandbox is ready, the trajectory's first model request joins the ready queue; this queue-entry event defines a prefill request arrival.
Admission to \textsc{Prefill} marks the start of service rather than a new arrival, and completion of prefill creates a decode admission and moves the trajectory to \textsc{Decode}.

The decoded response either terminates the trajectory or contains a tool call.
In the latter case, the trajectory enters \textsc{Tool} and waits for the external tool or environment to return.
The tool result is appended to the trajectory context, creating a new prefill request whose arrival is likewise recorded when it joins the ready queue.
If the decoded response reaches a stopping condition, the trajectory enters \textsc{Reward} and contributes no further requests.

\subsubsection{Workload Prediction}
\label{sec:design-prediction-procedure}

With the trajectory state machine, \sysname can predict the prefill workload and decode-admission token volume from four sources: the data sampler, sandbox startup, the ready queue, and tool execution.
The following summarizes each component and how it combines with in-flight work.
All calculations incur negligible overhead.
Due to space constraints, we leave the problem formulation and detailed equations to Appendix~\ref{sec:appendix-prediction}.

\begin{itemize}[leftmargin=*]
    \item \textbf{Data sampler.}
    The sampler estimates how many trajectories currently in the reward phase will complete within the horizon.
    It applies an estimated training-admission rate (the fraction of reward-completed trajectories meeting training-admission criteria) and the sample-injection policy to infer new launches, then accounts for dispatch and startup delay before converting predicted request arrivals into prefill demand using estimated prompt lengths.
    Newly sampled requests are assumed to hit the prewarm pool, but their first requests need not arrive immediately.

    \item \textbf{Sandbox startup.}
    For sandboxes already starting, the predictor distinguishes prewarm hits from cold starts.
    For both modes, historical startup-to-request durations, conditioned on elapsed time, determine the probability that the first request reaches the ready queue within the horizon.

    \item \textbf{Ready queue.}
    All requests already waiting in the ready queue contribute their full prompt-token counts as observed backlog.

    \item \textbf{Tool execution.}
    For trajectories currently executing tools, the predictor uses historical tool durations and elapsed execution time to estimate when the next model request will arrive.
    Expected arrivals within the horizon contribute the estimated prompt length after the tool result is appended.

    \item \textbf{In-flight prefill requests.}
    All requests already in \textsc{Prefill} contribute their full prompt-token counts.

    \item \textbf{Decode admissions from prefill completion.}
    The predictor estimates prefill completion times from runtime progress, request arrival times, per-instance queue order, and profiled execution times.
    Requests expected to complete prefill within the horizon contribute their prompt lengths to the decode-admission token forecast.
    These token counts are converted to KV-cache requirements under the selected decode configuration, without predicting output lengths.

    \item \textbf{Aggregation and online correction.}
    The predictor sums the prefill contributions and refreshes the forecasts every $S$ seconds.
    Observed arrivals and completions update duration statistics, training-admission outcomes update the acceptance-rate estimate, and forecast residuals drive an online bias correction.
\end{itemize}

\subsection{Prefill-Decode Role Switching}
\label{sec:design-switch}

We select the parallelism configuration with the highest measured throughput separately for prefill and decode.
For workloads with particularly long requests, a larger decode EP size is used to provide more KV-cache capacity.
In agentic RL, KV-cache capacity limits concurrency for long requests, making it a key constraint on decode throughput.
Therefore, the role switcher uses decode KV-cache requirements combined with prefill demand to set the PD instance ratio.

At time $t$, let $\mathcal{W}^{\mathrm P}_t$ denote the estimated full-prompt token demand for allocation over a prediction horizon of duration $S$.
It includes all queued and in-flight prefill requests at $t$ and new requests predicted to arrive within the horizon, rather than only requests expected to complete within it.
$K_t^{\mathrm D}$ is the current decode KV-cache footprint, and $\hat K^{\mathrm D}_{t,S}$ is the KV-cache space required by requests predicted to enter decode within the horizon.
The switcher estimates the effective prefill processing speed $\theta^{\mathrm P}_t$ in full-prompt tokens per second per instance, using the same counting convention as $\mathcal{W}^{\mathrm P}_t$, including cached prefixes.
This effective speed incorporates the workload's average prefix-cache hit ratio: the benefit of cache reuse is reflected in the measured service time, while token counts retain the full prompt lengths.
It initializes this estimate from profiling and updates it from recent busy intervals under the selected configuration with comparable prompt-length distributions and average prefix-cache hit ratios.
Each completed request contributes its full prompt length once; the denominator sums busy instance-seconds rather than wall-clock time, excluding intervals idle for lack of requests.
The online measurement uses uninterrupted, completed busy intervals, from the start of service on an idle instance until its prefill work drains, so token counts and service time cover the same requests; overlapping requests share elapsed instance time rather than being timed separately.
If no matching busy interval has completed, the switcher retains the last valid estimate for that workload profile, falling back to profiling when needed, rather than treating idle observations as zero capacity.
Let $C_{\mathrm{KV}}^{\mathrm D}$ denote the usable KV-cache capacity per decode instance.
The switcher computes the prefill instance count from throughput and the decode instance count solely from KV-cache capacity:
\begin{equation}
  n^{\mathrm P}_t=\Big\lceil\frac{\mathcal{W}^{\mathrm P}_t}{S\times\theta^{\mathrm P}_t}(1+\delta)\Big\rceil,
  \qquad
  n^{\mathrm D}_t=\Big\lceil\frac{K_t^{\mathrm D}+\hat K^{\mathrm D}_{t,S}}{C_{\mathrm{KV}}^{\mathrm D}}(1+\delta)\Big\rceil,
  \label{eq:switch-ratio}
\end{equation}
where $\delta\ge0$ is a shared safety margin.
Within our short prediction horizon, we do not explicitly model KV-cache growth during decoding or cache reclamation upon request completion, as these are hard to predict.
The safety margin $\delta$ provides KV-cache headroom to sustain decode throughput.

Let $m^{\mathrm P}$ and $m^{\mathrm D}$ be the numbers of GPUs per instance, and let $G$ be the total GPU budget.
Our objective is to maximize $\delta$ subject to the GPU budget constraint $m^{\mathrm P}n^{\mathrm P}_t+m^{\mathrm D}n^{\mathrm D}_t\le G$.
The resulting instance counts define the target PD ratio $n^{\mathrm P}_t:n^{\mathrm D}_t$.
If the predicted demands cannot be met within the budget even at $\delta=0$, the switcher instead chooses a budget-feasible integer allocation that minimizes the larger of the prefill throughput and decode KV-cache demand-to-capacity ratios.
Appendix~\ref{sec:appendix-switch} details the switch execution procedure.

\section{Evaluation}
\label{sec:evaluation}

We implement \sysname on top of Verl~\citep{sheng2025hybridflow} and SGLang~\citep{sglang}.
Our modifications comprise around 6.4K lines of Python code and support scheduling and role switching.
In this section, we evaluate \sysname
in terms of overall performance (\S\ref{sec:perf-overall}),
sources of performance gains (\S\ref{sec:perf-breakdown}),
role-switching effectiveness (\S\ref{sec:perf-ablation}),
and workload prediction accuracy (\S\ref{sec:perf-prediction}).

\subsection{Experimental Setup}
\label{sec:eval-setup}

\begin{figure}[t]
\centering
\includegraphics[width=0.95\linewidth]{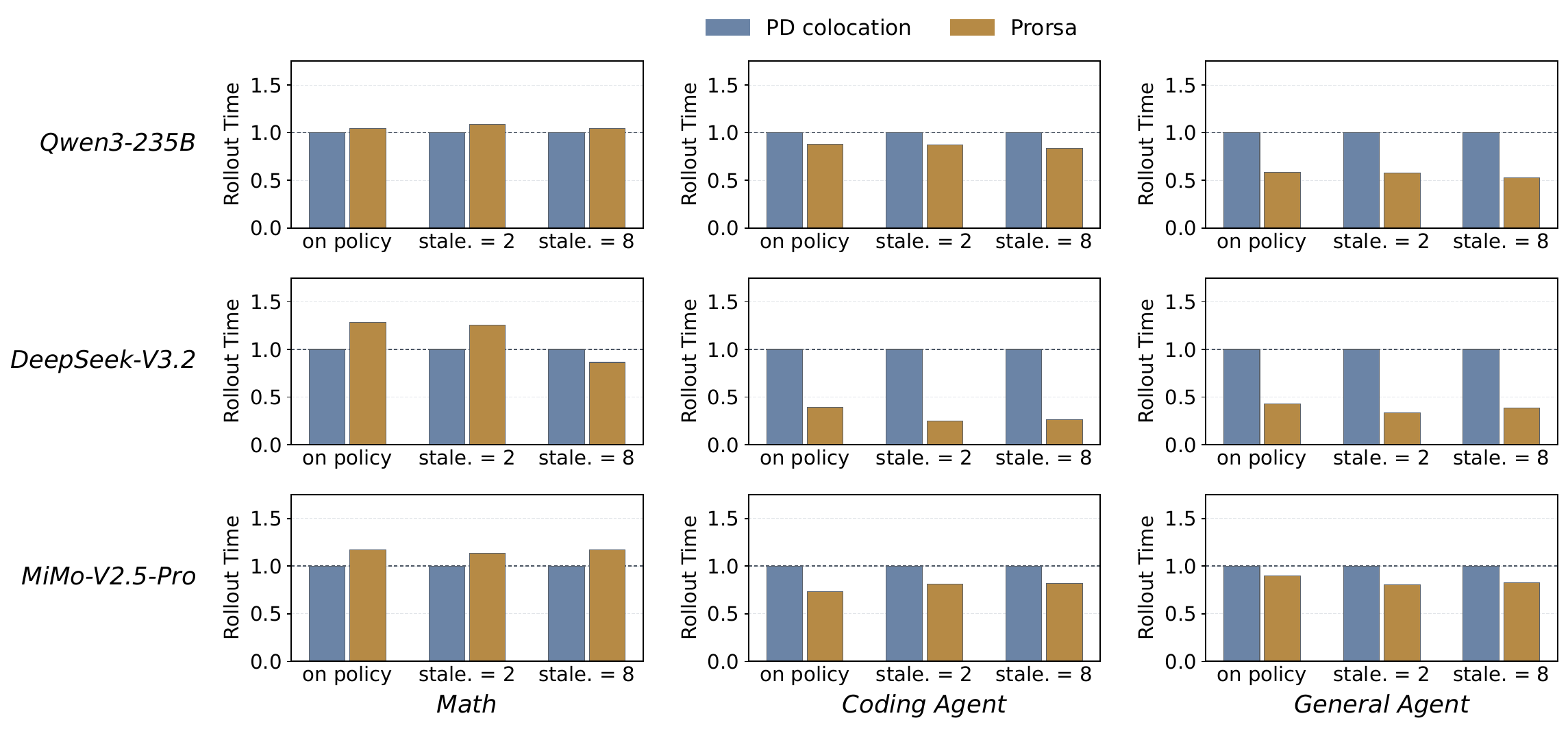}
\caption{Overall performance of \sysname and PD colocation.}
\label{fig:perf-overall}
\vspace{-1em}
\end{figure}

\textbf{Testbed.} Experiments are conducted on a GPU cluster with 16 servers,
each with 8 GPUs connected via NVLink. Each server is equipped with eight 400 Gbps RDMA NICs for inter-server connectivity.

\textbf{Models.} We use three models: (a) Qwen3-235B-A22B~\citep{qwen3technicalreport}, (b) DeepSeek-V3.2 (671B)~\citep{deepseekai2025deepseekv32}, and (c) MiMo-V2.5-Pro (1T)~\citep{mimo2026v25pro}.
We enable chunked prefill~\citep{sarathi} and use SGLang's recommended chunk sizes (4096 for Qwen3-235B and MiMo-V2.5-Pro, and 512 for DeepSeek-V3.2).
All models use FP8 quantization. For each model, we select an optimal parallelism configuration that allows the model to fit within GPU memory while reserving sufficient KV-cache space for agentic RL rollout.
We set the expert-parallel size to 16 for DeepSeek-V3.2 and MiMo-V2.5-Pro, and to 4 for Qwen3-235B.
For attention tensor parallelism, Qwen3-235B uses TP=2 and MiMo-V2.5-Pro uses TP=8.
For DeepSeek-V3.2, we disable attention tensor parallelism since it would duplicate the KV cache across TP ranks.

\textbf{Workloads.} We use the following three representative RL tasks.
\begin{itemize}[leftmargin=*]
\item \emph{Math}: single-turn mathematical problem solving tasks from AIME25~\citep{aime25}.
\item \emph{Coding Agent}: software engineering tasks from SWE-smith~\citep{yang2025swesmith}.
\item \emph{General Agent}: general-purpose tasks from APEX-Agents~\citep{vidgen2026apexagents}, Toolathlon~\citep{li2026tooldecathlonbenchmarkinglanguage}, and MCP-Atlas~\citep{bandi2026mcpatlaslargescalebenchmarktooluse}.
\end{itemize}
All tasks use the GRPO algorithm~\citep{shao2024deepseekmath}, with a group size of 8 for the math task and 32 for the two agentic tasks.
We use a 10-second workload prediction window in all experiments.

\subsection{Overall Performance}
\label{sec:perf-overall}

Figure~\ref{fig:perf-overall} compares the average rollout times of \sysname and PD colocation across different settings.
The results indicate that \sysname reduces end-to-end rollout time on the evaluated agentic tasks.
For the two agentic RL tasks, \sysname delivers a 1.11$\times$--4.05$\times$ speedup across different staleness settings, demonstrating the potential of PD disaggregation.
Specifically, for DeepSeek-V3.2, which has 16 individual attention modules per engine, \sysname improves rollout efficiency by 2.54$\times$--4.05$\times$ on the coding agent task and 2.35$\times$--2.99$\times$ on the general agent task.
We attribute most of this speedup to reduced interference among attention modules, as detailed in \S\ref{sec:perf-breakdown}.
In contrast, on the math task, normalized rollout step times range from 0.86 to 1.28 relative to colocation, indicating a slowdown in most evaluated settings.
This is due to much less overlap between prefill and decode: rollout first prefills all sequences and then executes successive decoding steps.
For these workloads, PD disaggregation incurs KV-cache transfer and role-switching overheads that degrade performance.

\subsection{Performance Breakdown}
\label{sec:perf-breakdown}

\begin{table}[t]
  \centering
  \caption{Average end-to-end trajectory completion time (seconds) on the general agent task.}
  \label{tab:general-agent-traj-time}
  \footnotesize
  \setlength{\tabcolsep}{2pt}
  \begin{tabular*}{\linewidth}{@{\extracolsep{\fill}}l*{9}{c}@{}}
    \toprule
    Model & \multicolumn{3}{c}{Qwen3-235B} & \multicolumn{3}{c}{DeepSeek-V3.2} & \multicolumn{3}{c}{MiMo-V2.5-Pro} \\
    \cmidrule(lr){2-4}\cmidrule(lr){5-7}\cmidrule(lr){8-10}
    Staleness & On-policy & 2 & 8 & On-policy & 2 & 8 & On-policy & 2 & 8 \\
    \midrule
    PD colocation & 133.88 & 169.14 & 176.55 & 713.05 & 723.96 & 939.54 & 253.17 & 198.08 & 252.36 \\
    \midrule
    \multirow{2}{*}{\sysname} & 78.98 & 86.96 & 80.40 & 119.47 & 123.21 & 128.12 & 195.35 & 158.90 & 216.42 \\
    & (1.7$\times\downarrow$) & (1.9$\times\downarrow$) & (2.2$\times\downarrow$) & (6.0$\times\downarrow$) & (5.9$\times\downarrow$) & (7.3$\times\downarrow$) & (1.3$\times\downarrow$) & (1.2$\times\downarrow$) & (1.2$\times\downarrow$) \\
    \bottomrule
  \end{tabular*}
  \vspace{-1em}
\end{table}

\begin{figure}[t]
  \centering
  \begin{minipage}[t]{0.5821\linewidth}%
    \centering
    \includegraphics[width=\linewidth]{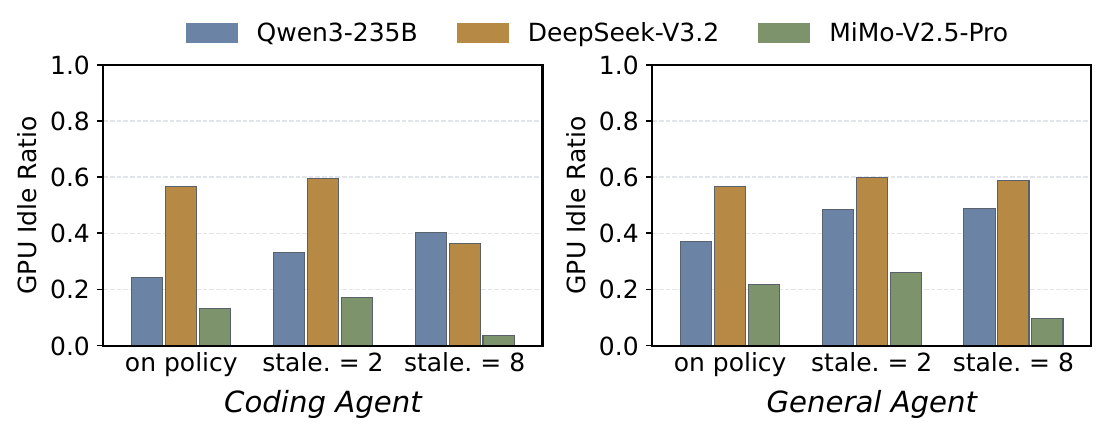}%
    \captionof{figure}{Average GPU idle ratio due to prefill-decode interference under PD colocation.}
    \label{fig:idle-ratio-overall}%
  \end{minipage}\hfill
  \begin{minipage}[t]{0.3979\linewidth}%
    \centering
    \includegraphics[width=\linewidth]{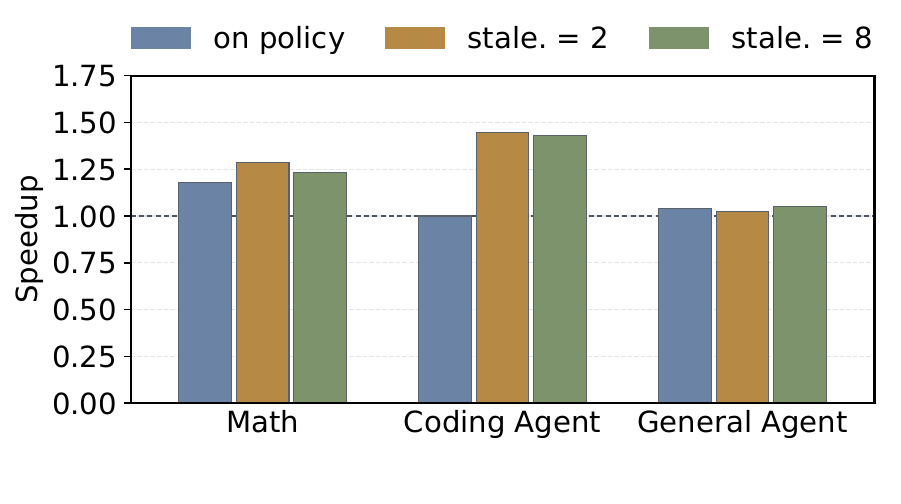}%
    \captionof{figure}{Effects of prediction-based role switching for MiMo-V2.5-Pro.}
    \label{fig:slowdown-overall}%
  \end{minipage}
  \vspace{-1em}
\end{figure}

\textbf{Improved GPU utilization.}
Figure~\ref{fig:idle-ratio-overall} reports an average GPU idle ratio of 3.76\%--59.98\% under PD colocation due to the prefill-decode interference across attention DP ranks.
For DeepSeek-V3.2 (DP=16), this interference is the dominant throughput bottleneck: a single prefill request can stall all other decode batches, yielding idle ratios of 36.21\%--59.98\%.
This idleness is the primary contributor to rollout slowdown.
Note that the reported idle ratios exclude GPU idleness arising from scheduling overhead and cross-DP prefill imbalance, whose contributions are hard to quantify accurately.
Therefore, these ratios cannot be directly translated into an end-to-end slowdown factor.

\textbf{Stage-specific parallelism.}
PD disaggregation enables independent configuration of prefill and decode parallelism to maximize throughput.
For the coding agent task, we double the expert-parallel size of decode instances for MiMo-V2.5-Pro and Qwen3-235B, yielding speedups of 1.04$\times$--1.09$\times$.

\textbf{Reduced decode latency.}
Table~\ref{tab:general-agent-traj-time} summarizes average trajectory completion times on the general agent task across different settings.
Compared with PD colocation, \sysname reduces end-to-end trajectory latency by 1.17$\times$--7.33$\times$.
This improvement arises because PD disaggregation mitigates both types of prefill-decode interference: (a) cross-DP imbalance that blocks decode batches, as discussed earlier, and (b) head-of-line blocking caused by higher-priority prefill requests.

\subsection{Ablation Study}
\label{sec:perf-ablation}

Figure~\ref{fig:slowdown-overall} evaluates the benefits of role switching by comparing \sysname with static PD disaggregation for MiMo-V2.5-Pro.
The static baseline uses the optimal prefill-to-decode instance ratios of 4:4, 5:3, and 3:5 for the math, coding agent, and general agent tasks, respectively.
The results show that role switching yields speedups of 1.0$\times$--1.44$\times$ across different settings.
On the math task, an initial prefill phase for all samples is followed by prolonged decoding.
A static PD ratio therefore underutilizes prefill and decode resources at different points in rollout.

The benefits of role switching depend on workload characteristics.
Coding agent trajectories have an average end-to-end completion time of 846 seconds, 3.92$\times$ that of general agent trajectories (216 seconds).
Longer trajectory lifetimes increase request concurrency, and asynchronous rollout further amplifies this effect through sample over-dispatch, placing greater pressure on decode KV-cache capacity and limiting throughput.
Role switching alleviates this capacity bottleneck by increasing the number of decode instances when needed.
In contrast, the general agent task benefits less from role switching due to lower KV-cache pressure.

\subsection{Prediction Accuracy}
\label{sec:perf-prediction}

\begin{figure}[t]
  \centering
  \begin{minipage}[t]{0.32\linewidth}%
    \centering
    \includegraphics[width=\linewidth]{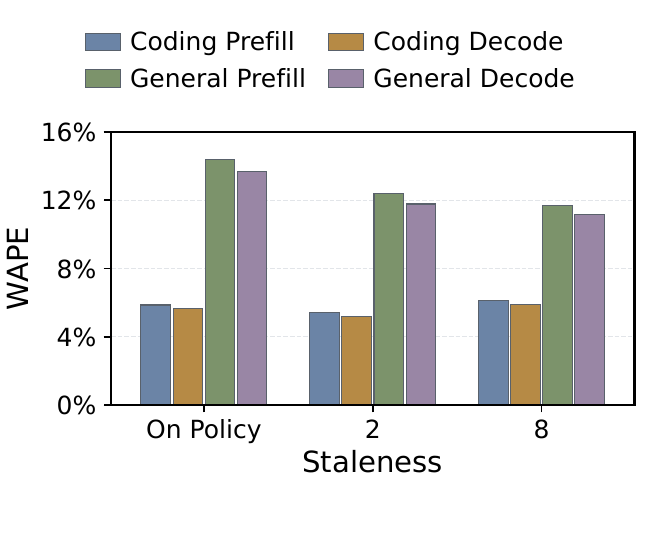}%
  \vspace{-1.5em}
    \caption{Prediction WAPE for MiMo-V2.5-Pro across staleness settings.}
    \label{fig:prediction-wape}%
  \end{minipage}\hfill
  \begin{minipage}[t]{0.64\linewidth}%
    \centering
    \includegraphics[width=\linewidth]{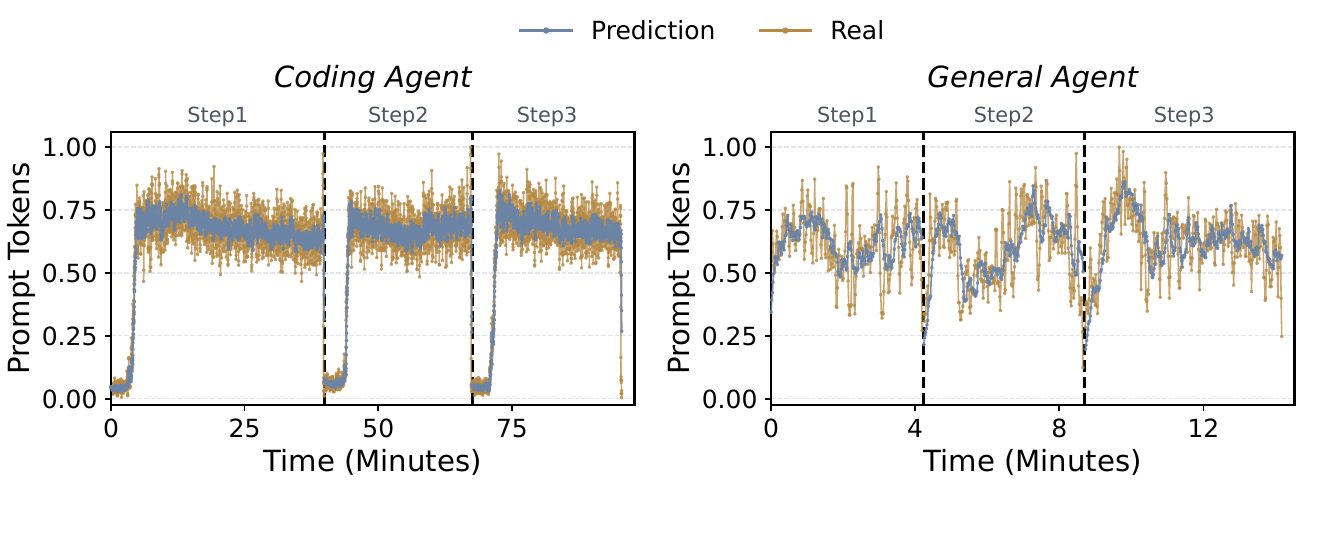}%
  \vspace{-1.5em}
    \caption{Predicted and measured normalized prefill workloads for MiMo-V2.5-Pro on the coding and general agent tasks. Dashed lines mark training-step boundaries.}
    \label{fig:prediction-accuracy}%
  \end{minipage}
  \vspace{-1em}
\end{figure}

Figure~\ref{fig:prediction-wape} reports prediction accuracy for MiMo-V2.5-Pro on the coding and general agent tasks across different staleness settings.
We evaluate prefill and decode-admission forecasts separately using weighted absolute percentage error (WAPE), defined as the sum of absolute prediction errors divided by the sum of measured token counts across all windows.
Across all settings, WAPE ranges from 5.40\% to 14.37\% for prefill and from 5.19\% to 13.69\% for decode.
This level of accuracy is sufficient to guide role switching.
The higher prediction errors on the general agent task reflect greater variability in tool-execution times.

Figure~\ref{fig:prediction-accuracy} further compares the prefill workload forecasts of \sysname with the measured workloads over the first three training steps with staleness set to 8.
The results show that, by monitoring all trajectories, \sysname produces forecasts that align with the measured workloads.

\section{Related Work}
\label{sec:related}

\textbf{RL frameworks.}
RL frameworks improve performance through resource and trajectory scheduling~\citep{sheng2025hybridflow, hu2024openrlhf, wang2025ROLL, wang2025distflow, wang2025seamlessflow, dora, gao2025rollpacker, zhong2024rlhfuse, mei2024real, wu2025weave}, asynchronous rollout~\citep{fu2025areal, zhong2025streamrl, han2025asyncflow, zhou2025aprilactivepartialrollouts}, and speculative decoding~\citep{he2025history, qin2025seer, chen2025respec, xu2026warworkloadawarerolloutssynchronous}.
Systems also adapt rollout parallelism~\citep{gao2025rollpacker, tan2026orchestrrldynamiccomputenetwork,heddle} or multiplex prefill and decode in place~\citep{ainfer-pd},
and a few production RL systems already adopt PD disaggregation for agentic RL rollout~\citep{glm5, minimaxm2, rollart} without in-depth analysis.
Our work characterizes the benefits and deployment challenges of PD disaggregation for agentic RL, and introduces role switching to adapt the PD ratio to changing rollout demand.

\textbf{Prefill-decode disaggregation.}
PD disaggregation separates prefill and decode to avoid interference and enable stage-specific configuration~\citep{distserve, splitwise, mooncake, distattention, hong2025semi, feng2025windserve, zhu2025nanoflow, forys2026doesdisaggregationpaysimulating, wang2026pddunleashingeconomicalflexible}.
Variants include hybrid aggregation and disaggregation~\citep{wang2025prefilldecodeaggregationdisaggregationunifying}, intra-GPU multiplexing~\citep{cui2025pdmux,duetserve, xu2026crossflowprefilldecodeelasticityagentic}, refined splits for multi-turn prefills~\citep{li2026prefillsequalppddisaggregation}, adaptive multi-round prefill scheduling~\citep{he2026efficientmultiroundllminference}, and load-driven PD autoscaling~\citep{dopd}.
In contrast to these serving systems, \sysname applies PD disaggregation to agentic RL rollout and uses workload predictions based on RL lifecycle signals.

\textbf{LLM workload prediction.}
Existing LLM serving systems predict workloads for provisioning, scheduling, and model prewarming~\citep{preserve, pillm, oserve, sageserve, tokenscale, warmserve, cachewise, continuum}.
RL frameworks also exploit workload estimates: OrchestrRL~\citep{tan2026orchestrrldynamiccomputenetwork} forecasts remaining rollout workloads, while Seer~\citep{qin2025seer}, RhymeRL~\citep{he2025history}, and RollPacker~\citep{gao2025rollpacker} use online probes or historical workload statistics to guide scheduling and resource allocation.
For agentic RL, Heddle~\citep{heddle} predicts trajectory completion times, while Libra~\citep{chen2026libraefficientresourcemanagement} uses tool-return signals to estimate remaining work and guide scheduling.
In contrast, \sysname combines signals from reward-driven sample injection, sandbox startup, and tool returns to forecast short-horizon, turn-level prefill and decode demand in agentic RL rollout. %
\section{Conclusion}
\label{sec:conclusion}

Agentic RL introduces continuous prefill requests, making PD disaggregation beneficial for the rollout phase.
However, our experiments show that a static prefill-decode ratio leads to resource under-utilization due to dynamic rollout workloads.
\sysname employs an adaptive PD role-switching strategy to accommodate fluctuating demand, guided by workload forecasts derived from trajectory lifecycle signals.
Our evaluation demonstrates that \sysname effectively unlocks the performance potential of PD disaggregation during rollout, achieving up to 4.05$\times$ the throughput of PD colocation.
{}

\newpage
\begingroup
\ULforem
\bibliography{xin}
\endgroup

\appendix
\onecolumn

\section{Additional Motivation Experiments}
\label{sec:appendix-motivation}

We extend the workload analysis in \S\ref{sec:challenge} with the complete plots of inter-step and intra-step workload variation for the two agentic tasks under different configurations.

\begin{figure}[htbp]
  \centering
  \includegraphics[width=\linewidth]{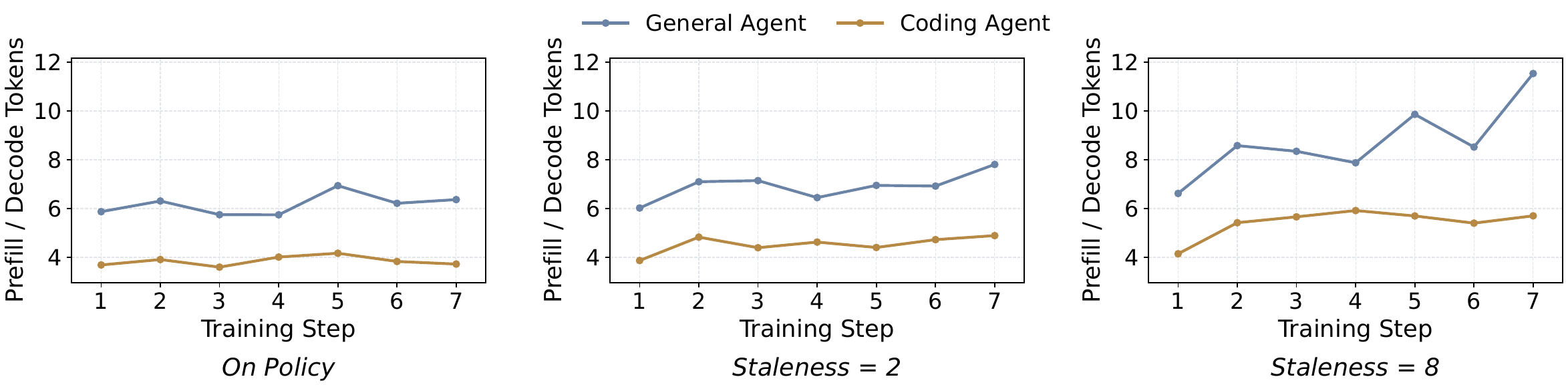}
  \caption{Inter-step prefill-to-decode token ratios for the general agent and coding agent tasks over seven training steps.}
  \label{fig:appendix-inter-step}
\end{figure}

\textbf{Inter-step variation.}
Figure~\ref{fig:appendix-inter-step} shows that workload variation occurs across all settings.
Even in the relatively stable on-policy setting, the maximum ratio is 1.20$\times$ the minimum.

\begin{figure}[htbp]
  \centering
  \includegraphics[width=\linewidth]{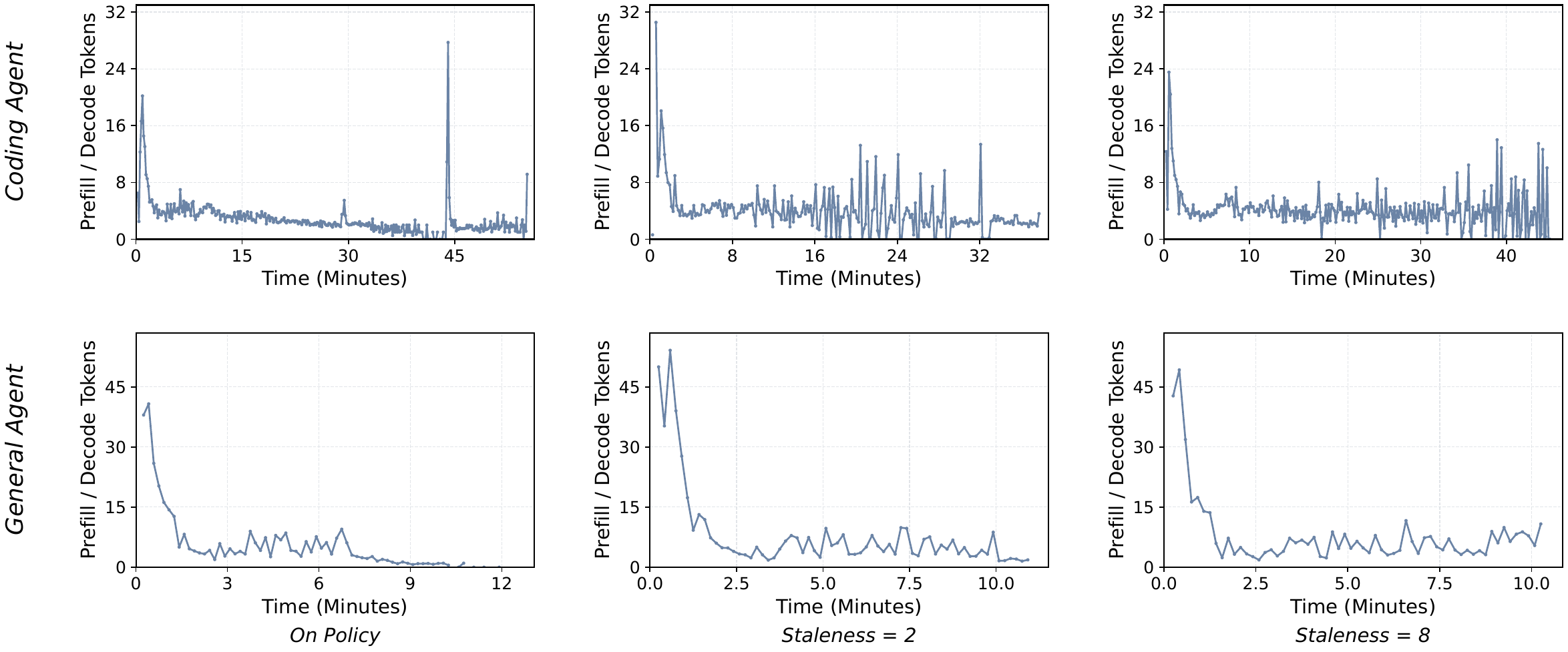}
  \caption{Intra-step prefill-to-decode token ratios within the first training step, measured over 10-second windows.}
  \label{fig:appendix-intra-step}
\end{figure}

\textbf{Intra-step variation.}
Figure~\ref{fig:appendix-intra-step} reveals a similar pattern of workload variation across all settings.
Across all settings, the first rollout step always begins with a prefill-heavy phase, followed by a sharp decline in the token ratio.

\section{Workload Prediction}
\label{sec:appendix-prediction}

This section formalizes the workload predictor in \S\ref{sec:design-prediction} and details the procedure summarized in \S\ref{sec:design-prediction-procedure}.
The predictor estimates prefill demand and decode-admission token volume from the data sampler, sandbox startup, the ready queue, tool execution, and observed in-flight requests.

\subsection{Problem Formulation}
\label{sec:appendix-prediction-formulation}

At time $t$, the predictor estimates two demand signals over the short horizon $[t,t+S)$ (default $S=10$\,s): a prefill workload estimate for worker allocation and the prompt-token volume of requests expected to be newly admitted to decode.
Throughout this section, $x_j$ denotes the observed or predicted full prompt length of request $j$, including any cached prefix; we do not subtract cached or already processed tokens from this length.
The processing-speed estimate in \S\ref{sec:design-switch} uses the same full-prompt convention and already incorporates the average prefix-cache hit ratio, so no separate cache-hit discount is applied to the demand.
Prefill demand is not a forecast of completed token volume.
For simplicity, we do not model a trajectory entering the prefill and decode stages more than once within the prediction horizon.

\textbf{Input.}
The input is an atomic snapshot of the rollout framework at time $t$, including the lifecycle record $T_t$ of every live trajectory: its current state, context length, elapsed time in that state, and group membership under group-based algorithms.
The snapshot further contains the online training-admission rate estimate $\hat p_t$, the configured sample-injection function $f(\cdot)$, the sandbox records $P_t$ for launched trajectories whose first requests have not yet joined the ready queue, the ready queues and their instance assignments, the in-flight prefill set $Q_t^P$ with each request's prompt length and instance assignment, the active decode set $Q_t^D$ with its current KV-cache footprint $K_t^{\mathrm D}$, and historical duration statistics for sandbox startup, tool execution, and reward computation.
Runtime finish-time estimates for in-flight prefills are additionally used to predict decode admissions, not to compute prefill demand.

\textbf{Output.}
The predictor outputs the prefill demand forecast $\hat W^{\mathrm P}_{t,S}$ and the decode-admission token forecast $\hat A^{\mathrm D}_{t,S}$ over $[t,t+S)$.
$\hat W^{\mathrm P}_{t,S}$ is measured in prompt tokens and follows the source-specific counting rules in \S\ref{sec:design-prediction-procedure}.
New requests predicted to arrive from the data sampler, sandbox startup, or tool execution before $t+S$ contribute their estimated prompt-token counts.
All requests already in the ready queue or $Q_t^P$ at $t$ contribute their full prompt lengths, regardless of their predicted completion times.
The forecast measures prompt-token volume rather than remaining token-processing work.
$\hat A^{\mathrm D}_{t,S}$ aggregates the prompt lengths of requests whose prefill is expected to complete before $t+S$.
The allocator combines the prefill forecast with the current decode footprint $K_t^{\mathrm D}$ and the KV-cache demand $\hat K^{\mathrm D}_{t,S}$ derived from predicted admissions to select worker roles in \S\ref{sec:design-switch}.

\subsection{Data Sampler}

In asynchronous rollout, reward completion triggers sample injection, so the sampler forecasts how many trajectories will complete reward computation within $[t,t+S)$.
Let $R_t$ be the set of trajectories currently in the reward phase.
For $i\in R_t$, the probability of completing reward computation within the horizon is
\begin{equation}
  q_i^{\mathrm{rew}}=\Pr(T_i^{\mathrm{rew}}<e_i+S\mid T_i^{\mathrm{rew}}>e_i),
  \label{eq:reward-finish}
\end{equation}
where $e_i$ is the elapsed reward time and $T_i^{\mathrm{rew}}$ follows a task-conditioned empirical distribution of total reward durations.
The conditioning accounts for the fact that reward computation has not finished by time $t$.
Here, a trajectory \emph{passes} if it satisfies the criteria for admission to training.
The online estimate $\hat p_t\in[0,1]$ approximates the fraction of reward-completed trajectories that satisfy these criteria.
Using this shared acceptance-rate estimate, we predict the number of newly eligible trajectories as
\begin{equation}
  N^{\mathrm{pass}}_{t,S}=\hat p_t\sum_{i\in R_t}q_i^{\mathrm{rew}}.
  \label{eq:pass-count}
\end{equation}
When all reward-completed trajectories are eligible for training, $\hat p_t=1$.
Under GRPO~\citep{shao2024deepseekmath}, some group members may have completed reward computation by time $t$.
The sum includes only members still in the reward phase, weighted by their probabilities of completing reward computation within the horizon, and is scaled by $\hat p_t$.
Members still in \textsc{Decode} or \textsc{Tool} are excluded, consistent with the one-step assumption.

Rollout systems inject new samples under different policies---replace-on-reward, fixed-rate refresh, or group replenishment---so we abstract the injection rule as a configurable function $f(\cdot)$.
The function is evaluated with the current sampler state, including known group-completion information, held fixed.
The estimated number of new sample launches triggered within the horizon is
\begin{equation}
  \hat N^{\mathrm{samp}}_{t,S}=f\!\left(N^{\mathrm{pass}}_{t,S}\right).
  \label{eq:sampler}
\end{equation}
For nonlinear or group-based injection rules, this is a plug-in approximation rather than an exact expectation over all possible completion outcomes.
For newly sampled requests, the forecast assumes a prewarm hit and omits cold starts of not-yet-launched samples.
Let $\hat d_t^{\mathrm{pw}}$ be the online mean delay from a sample-injection decision until the first request joins the ready queue, including dispatch and prewarmed startup.
Using this point estimate, define the effective launch horizon $S_t^{\mathrm{pw}}=\max(0,S-\hat d_t^{\mathrm{pw}})$ and reevaluate Equations~\ref{eq:reward-finish}--\ref{eq:sampler} over this shorter horizon, with the same sampler state and injection policy.
Each launch predicted early enough for its first request to arrive before $t+S$ contributes the estimated prompt-token count $\hat x$:
\begin{equation}
  \hat W^{\mathrm P,\mathrm{samp}}_{t,S}
    =\hat N^{\mathrm{samp}}_{t,S_t^{\mathrm{pw}}}\,\hat x.
  \label{eq:sampler-work}
\end{equation}
For $S_t^{\mathrm{pw}}=0$, the contribution is zero.
Thus, a launch within $[t,t+S)$ is not counted as a prefill arrival in that window if its estimated dispatch and startup delay extends beyond the horizon.

\subsection{Sandbox Startup}

For each sandbox $k$ already starting, let $a_k$ be its launch time and $\mathrm{mode}(k)\in\{\mathrm{pw},\mathrm{cold}\}$ identify a prewarm hit or cold start.
The first-request arrival time is $a_k+T_k^{\mathrm{start}}$, where $T_k^{\mathrm{start}}$ follows the measured launch-to-ready-queue duration distribution for that startup mode, including dispatch of the first request.
Since the first request has not arrived at time $t$, let $e_k=t-a_k$ and condition on this duration exceeding $e_k$:
\begin{equation}
  q_k^{\mathrm{start}}=\Pr(T_k^{\mathrm{start}}<e_k+S\mid T_k^{\mathrm{start}}>e_k,\mathrm{mode}(k)).
  \label{eq:cold-finish}
\end{equation}
With $P_t^{\mathrm{pw}}$ and $P_t^{\mathrm{cold}}$ denoting the sets of prewarm-hit and cold-start sandboxes in $P_t$, the expected contribution is
\begin{equation}
  \hat W^{\mathrm P,\mathrm{start}}_{t,S}
    =\sum_{k\in P_t^{\mathrm{pw}}} q_k^{\mathrm{start}}\,x_k
    +\sum_{k\in P_t^{\mathrm{cold}}}
      q_k^{\mathrm{start}}\,x_k.
  \label{eq:sandbox-work}
\end{equation}
Startup histories are conditioned on startup mode, image, node class, and concurrent startup load.

\subsection{Ready Queue}

Let $\mathrm{Ready}_t$ denote all requests waiting to start prefill in the snapshot at time $t$.
The ready-queue contribution is the full prompt-token volume of this observed backlog:
\begin{equation}
  \hat W^{\mathrm P,\mathrm{ready}}_{t,S}
    =\sum_{j\in\mathrm{Ready}_t} x_j.
  \label{eq:ready-work}
\end{equation}
Predicted future arrivals are accounted for by the sampler, sandbox-startup, and tool-execution components and are not counted again in $\mathrm{Ready}_t$.

\subsection{Tool Execution}

We forecast tool-induced prefill arrivals only for trajectories already in \textsc{Tool}.
For trajectory $i$, let $\tau_i$ be the exponential moving average (EMA) of its tool-execution duration and $\hat\tau_i$ the most recently observed duration in the same conditioning class; the update is
\begin{equation}
  \tau_i\leftarrow(1-\alpha)\tau_i+\alpha\hat\tau_i,
  \qquad \alpha=0.1.
  \label{eq:ema}
\end{equation}
The EMA is conditioned on task type, tool type, turn index, and concurrent tool load whenever enough samples are available.
Let $e_i$ be the elapsed time in the current \textsc{Tool} state; the estimated residual time to the next prefill arrival is
\begin{equation}
  r_i=\max(0,\tau_i-e_i).
  \label{eq:residual}
\end{equation}
Only trajectories whose tools are predicted to finish within the horizon, i.e., $r_i<S$, contribute, so the predicted tool-induced prefill workload is
\begin{equation}
  \hat W^{\mathrm P,\mathrm{tool}}_{t,S}
    =\sum_{i\in\mathrm{Tool},\,r_i<S}\hat x_i,
  \label{eq:tool-work}
\end{equation}
where $\hat x_i$ is the estimated prompt length after the tool result is appended.

\subsection{In-flight Prefill Requests}

Requests already in \textsc{Prefill} are observed directly.
Their contribution is the full prompt-token volume of all requests in $Q_t^P$:
\begin{equation}
  W^{\mathrm P,\mathrm{inflight}}_t
    =\sum_{j\in Q_t^P}x_j.
  \label{eq:prefill-drain}
\end{equation}
Assuming individual prefill executions are short relative to the prediction horizon, we approximate in-flight demand using full prompt lengths rather than explicitly subtracting already processed tokens.

\subsection{Decode Admissions from Prefill Completion}

We estimate the KV-cache footprint of newly admitted requests from their prompt lengths, without predicting output lengths.
Let $R^P_{t,S}$ contain the in-flight prefills, requests already in the ready queue, and the one-step future arrivals represented in the forecast.
Let $\hat b_j$ denote the predicted prefill completion time of request $j$, used only for decode-admission prediction.
For requests already in $Q_t^P$, this estimate comes from the runtime snapshot and accounts for execution progress before $t$; their completion times are not recomputed as if prefill were starting from scratch.

For queued requests and predicted future arrivals, let $g(x_j)$ denote the profiled prefill execution time for prompt length $x_j$ under the selected configuration.
We estimate completion times using a request-level serial approximation within each prefill instance, while modeling different instances in parallel.
Routing and queue order follow the current scheduling policy; the approximation does not explicitly simulate batched or chunked execution.
Let $u_j$ be the observed or predicted request-arrival time, with $u_j=t$ for requests already in the ready queue.
For future arrivals, $u_j$ includes the estimated residual reward or sandbox-startup delay, any subsequent sample-injection and prewarmed-startup delay, or the tool-return delay, as applicable.
Residual reward and sandbox-startup estimates are conditioned on the operation still being unfinished at $t$, as in the preceding subsections.
Let $v_m$ denote the next available time of prefill instance $m$, initialized to the latest estimated finish time of its in-flight requests, or $t$ if it is idle.
For each not-yet-started request $j$ assigned to instance $m(j)$, the estimated prefill finish time is
\begin{equation}
  \hat b_j=\max\{u_j,v_{m(j)}\}+g(x_j),
  \qquad v_{m(j)}\leftarrow\hat b_j.
  \label{eq:ready-delay}
\end{equation}
This recurrence is evaluated in scheduling order on each instance for queued requests and predicted future arrivals.
It prevents a future request from beginning prefill before it arrives and avoids serializing work assigned to different instances.

Let $\kappa^{\mathrm D}(x_j)$ denote the KV-cache footprint of a request with prompt length $x_j$ at decode admission under the selected decode configuration.
Only requests with $\hat b_j<t+S$ contribute to the admission forecasts:
\begin{equation}
  \begin{aligned}
    \hat A^{\mathrm D}_{t,S}
      &=\sum_{j\in R^P_{t,S},\,\hat b_j<t+S}x_j,\\
    \hat K^{\mathrm D}_{t,S}
      &=\sum_{j\in R^P_{t,S},\,\hat b_j<t+S}\kappa^{\mathrm D}(x_j).
  \end{aligned}
  \label{eq:decode-admission}
\end{equation}

\subsection{Aggregation and Online Correction}

The complete prefill forecast is
\begin{equation}
  \hat W^{\mathrm P}_{t,S}
   =\hat W^{\mathrm P,\mathrm{samp}}_{t,S}
    +\hat W^{\mathrm P,\mathrm{start}}_{t,S}
    +\hat W^{\mathrm P,\mathrm{ready}}_{t,S}
    +\hat W^{\mathrm P,\mathrm{tool}}_{t,S}
    +W^{\mathrm P,\mathrm{inflight}}_t.
  \label{eq:total-work}
\end{equation}
The forecast is refreshed every $S$ seconds using the current allocation.
Observed training-admission outcomes update $\hat p_t$; arrivals and completions update the reward and startup distributions and the tool-duration EMA, while forecast residuals update a multiplicative bias correction.

\section{Role Switch}
\label{sec:appendix-switch}

To implement the target allocation from \S\ref{sec:design-switch}, the executor selects the least-loaded eligible instance for each required role change.
The instance is removed from the routing pool and stops accepting new requests.
Affected requests are cancelled, with their input and already generated tokens retained and the KV cache state persisted to remote storage.
The rollout controller retries these requests on other instances using the retained prefixes and persisted KV cache, avoiding recomputation of cached tokens.
Retries continue generation from the tokens already produced rather than restarting from the original input, preserving the output already visible to the agent.
This cancel-and-resend mechanism is transparent to the agent sandbox and requires no intervention from it.
Once in-flight work has quiesced, the engine rebuilds role-specific communication and cache resources for the target role, coordinating the transition across all participating ranks.
Model weights remain resident in GPU memory throughout execution, so no additional weight loading is required during role switching.
The whole role-switch procedure takes around 5 seconds on average.

\end{document}